\documentclass{article}
\usepackage[utf8]{inputenc}
\usepackage[T1]{fontenc}
\usepackage{amsmath}
\usepackage{amsfonts}
\usepackage{latexsym}
\usepackage{amssymb}
\usepackage{times}
\usepackage{ifpdf}
\usepackage{makeidx}
\usepackage{graphicx}
\ifpdf
 \usepackage[pdftex,colorlinks]{hyperref}
\else
 \usepackage[ps2pdf,breaklinks=true,colorlinks=true,linkcolor=red,citecolor=green]{hyperref}
 \fi

\newcommand{\Q}{{\mathbb{Q}}}

\title{Computing huge Groebner basis like cyclic10 over $\Q$ with Giac.}

\author{Bernard Parisse\\Institut Fourier\\UMR 5582 du
  CNRS\\Universit\'e de Grenoble I}
\date{2019}

\begin{document}

\maketitle

\begin{abstract}
We present a short description on how to fine-tune the modular
algorithm implemented in the Giac computer algebra system
to reconstruct huge Groebner basis over $\Q$.
The classical cyclic10 benchmark will serve as example.
\end{abstract}

\section{Introduction}
In 1999, J.-C. Faug\`ere presented the F4 algorithm (\cite{F99a}),
it was illustrated with the reconstruction of the Groebner basis over
$\Q$ of cyclic9,
a symmetric Groebner basis with 9 variables, obtained in 18 days.
The cyclic9 basis has 1344 elements, 1067837 monomials and requires
378 29-bits primes for reconstruction.
Twenty years later, with more powerful computers (more RAM) and multi-CPU
architectures, the same computation takes about 20 minutes
with Giac/Xcas (\cite{giac})). The new big benchmark in this category is now cyclic10, with 10
variables instead of 9. The computation modulo a prime takes a
few hours with a generic algorithm (i.e. without taking in account the
symmetries like in \cite{steidel2013grobner} or \cite{faugere2013grobner}), 
the basis has 5690 elements, about 20 millions monomials 
(19825931 or a little less for some primes, because of
cancellation). The computation over $\Q$ is a challenge\footnote{as
  far as I know, this document reports the first successful attempt
  to compute cyclic10 over $\Q$. Allan Steel reported about cyclic9
  over $\Q$ in Magma 15 years ago \cite{magma}, 
more recent reports are in finite fields}, because most
coefficients require more than 2000 29-bits primes for reconstruction, and there are about 20 millions
of coefficients to compute, i.e. the Groebner basis itself is more than 50G of RAM.
Being able to solve this example is not intrinsically interesting 
(especially since our algorithm does not exploit any symmetries), 
but induced improvements in many areas in the Giac's modular gbasis
algorithm implementation. I also had
to implement partial computation saving.

Section 2 will briefly explain the different phases of the modular
reconstruction algorithm. Section 3 will describe how to control
Giac's \verb|gbasis| command and show how we were able
to compute the Groebner basis of cyclic10 on $\Q$
on a server with 256G of RAM, 36 CPU (of 48) in
about 10 days of real-time (more than 200 days of sequential time).

\section{Modular reconstruction algorithm}
The modular gbasis algorithm over $\Q$ is done in a loop with
three steps~: gbasis computation modulo a prime, Chinese remaindering
with previous primes and rational reconstruction attempts. 
Rational reconstruction is not checked on each basis element after each prime,
this would take too long. The basis elements are sorted by increasing revlex
order, then an attempt is made to reconstruct the first basis element that
is not yet reconstructed and check that it fits with the last prime.
If reconstruction is successful, try the next
one and so on until reconstruction fails. 
As a side effect, basis elements reconstruct in
clusters (it is not clear if there is a better ordering strategy of
basis elements to attempt reconstruction, I tried a few obvious
strategies like age, but it was worse than revlex order).
Already reconstructed generators are checked each time a new prime gbasis
is computed.

For multi-CPU architectures, there are two ways to parallelize the
computation~:
\begin{itemize}
\item {\bf multiple primes} \\
compute the Groebner basis modulo several primes in parallel. This gives the
best speed-up (about the same real-time/CPU time ratio than the number of simultaneous primes), 
but it requires more memory (for example with Giac, about 12G of
RAM per prime for cyclic10, running more than 12 primes simultaneously
is not a good idea with 256G of RAM).
\item {\bf 1-prime parallelism}\\
most tasks in a modular Groebner basis computation can be paralleled: sparse
linear algebra, Gaussian reduction, inter-reduction... This does not require additional
memory (vs mono-threaded execution), but is not as efficient as computing
multiple primes simultaneously.
\end{itemize}
Note that Chinese remaindering (and rational reconstruction) 
is single-threaded, because it can trigger memory
allocations (and locks between threads could slow down the real-time
computation). With a better knowledge of GMP memory handling, this could
probably be paralleled.

There are two classical optimizations ~:
\begin{itemize}
\item {\bf learning}\\
The idea is that the F4 algorithm will follow the same path for all
primes: the same pairs will reduce to 0 and the same monomials will
be involved (``symbolic preprocessing'' information), we can record
them for further runs. For more details, see \cite{parisse2013probabilistic}.
\item {\bf re-injection}\\
Some of the ideal generators have smaller coefficients than other,
reconstruction can happen after a fraction of the total number of required
primes (see also \cite{monagan2017algorithm}). Using this
additional information, we can speed up the computation for the
remaining primes. For example, partial reconstruction happens for cyclic10 after 390
29-bits primes (for 20 generators), then after 794 primes (for 2164
generators).
Using this additional information will also save memory 
because the gbasis algorithm requires less intermediate computations
and coefficients storage is optimized: for example a
400 bits reconstructed rational (with numerator and denominator of
size 200 bits) takes 3 times less memory than it's integer
representant modulo a 1200 bits integer.
\end{itemize}

Examples of observed real-time speeds computing cyclic10 gbasis modulo a 29
bits prime using at most 36 CPU (\verb|cocalc|: Intel Xeon CPU E5-2687W v4 @
3.00GHz, first part of the table) or at most 24 CPU (\verb|ifnode2|: Intel Xeon CPU E5-2640
v3 @ 2.60GHz, second part).
Run 1 means first prime run modulo a 29 bit prime (includes storing learning
information for further runs).
Run 2 means next primes runs modulo a 29 bits prime with learned
information.
Time does not include Chinese remaindering (varying from a few seconds
at stage 1 start to almost 1 minute per prime at stage 2 end) or
rational reconstruction
(negligible except when clusters of generators reconstruct)\\
\begin{tabular}{|c|c|c|c|c|} \hline
Run & Re-inject. & simult. primes & threads per prime & minutes per
                                                       prime \\ \hline
1 & No & 1 & 24 & 200 \\
2 & No & 4 & 6 & 21.25 \\
2 & No & 3 & 8 & 24 \\
2 & No & 2 & 12 & 50\footnote{Measured near the end of computation
                  stage 1 with 90\% memory}\\
1 & 2205 & 1 & 24 & 5 \\
2 & 2205 & 4 & 6 & 1 \\
\hline
1 & No & 1 & 36 & 180 \\
2 & No & 12 & 3 & 12.5 \\
2 & No & 9 & 4 & 15 \\
1 & 2205 & 1 & 36 & 5 \\
2 & 2205 & 9 & 4 & 0.5  \\
\hline
\end{tabular}\\
For comparison with a few other implementations, see for example
\cite{ederimplementing} (generic code w/o learning: about 1 hour)
and \cite{faugere2013grobner} (code using symmetries, without
learning, computes the gbasis for a 16-bits prime
in 5 minutes with slower hardware, our run 2 with generic code does
a 29-bits prime in 12.5 minutes).

Therefore the modular algorithm is done in several phases~:
\begin{itemize}
\item phase 0~: 1st prime run with parallelism, collect
information for further runs
\item phase 1~: next runs (several primes, for each prime 
parallelism by number of
  CPU available divided by number of simultaneous prime runs).
No re-injection at this stage yet. Partial reconstructions attempts. 
\item phase 2~: if at some point, partial reconstruction
of a significant number of generators
is realized in phase 1, re-inject reconstructed generators, clear learned
information, redo phase 0 with initial generators+reconstructed
generators, then redo phase 1 until final reconstruction, or partial
reconstruction of a larger part of the basis (phases 3 or more)
\item phase 3, 4, ...~\\
like phase 2 with a more complete partial reconstructed basis
\end{itemize}
For huge computations, it may be desirable to save partial
computations, this can be done here by archiving reconstructed
generators. Further computations can start at phase 2
instead of phase 0.

\section{Fine-tuning Giac gbasis}
\subsection{Fine-tuning commands}
In addition to the \verb|threads:=n| command to control
the number of threads, Giac 1.5.0-49 introduces some new instructions to control the 
\verb|gbasis| implementation~:
\begin{itemize}
\item \verb|gbasis_simult_primes(n)|\\
\verb|gbasis_simult_primes(n1,p1,n2,p2,n3)|\\
This parameter controls the number of simultaneous modular prime
computations.
With 5 arguments instead of 1, the number of simultaneous modular computation
is defined as \verb|n1| until \verb|p1| primes are computed, then
\verb|n2| until \verb|p2| primes, then \verb|n3|.
\\
The best real-time is obtained by choosing the highest number of primes
compatible with the RAM available, then divide the number of available
threads by this number of primes and give it for each 1-prime
parallelism.
\item \verb|gbasis_reinject(ratio,speed_ratio)|\\
Change defaults re-injection parameters. If the number of reconstructed
generators over the number of elements of the basis is greater than 
\verb|ratio| and the speed ratio of the second run (with learning
information) vs first run (without learning information) is greater
than \verb|speed_ratio|, then re-injection will happen.
\item \verb|gbasis_reinject(-n)|\\
With a negative integer as argument, this will stop the computation
as soon as at least $n$ elements of the Groebner basis are
reconstructed and \verb|gbasis| will return a partial Groebner basis.
This is used in conjunction with\\
\verb|archive("filename",gbasis(list_generators,vars))|\\
for a computation in several stages. Further stages will call\\
\verb|gbasis(list_generators,vars,gbasis_reinject=unarchive("filename"))|
\item \verb|gbasis_max_pairs(n)|\\
Maximal number of pairs that will be reduced simultaneously in the F4
algorithm. This can reduce time and memory for a 1-prime computation but it 
seems to be less efficient on further runs.
\end{itemize}

\subsection{Example with cyclic9}
The computation of cyclic9 can be done in 1 stage with the following script\\
{\tt \small
cyclic9:=[x1 + x2 + x3 + x4 + x5 + x6 + x7 + x8 + x9, 
x1*x2 + x2*x3 + x3*x4 + x4*x5 + x5*x6 + x6*x7 + x7*x8 + x1*x9 + x8*x9, x1*x2*x3 + x2*x3*x4 + 
x3*x4*x5 + x4*x5*x6 + x5*x6*x7 + x6*x7*x8 + x1*x2*x9 + x1*x8*x9 + x7*x8*x9, 
x1*x2*x3*x4 + x2*x3*x4*x5 + x3*x4*x5*x6 + x4*x5*x6*x7 + x5*x6*x7*x8 +
x1*x2*x3*x9 + x1*x2*x8*x9 + x1*x7*x8*x9 + x6*x7*x8*x9, x1*x2*x3*x4*x5
+ x2*x3*x4*x5*x6 + x3*x4*x5*x6*x7 + x4*x5*x6*x7*x8 + x1*x2*x3*x4*x9 + 
x1*x2*x3*x8*x9 + x1*x2*x7*x8*x9 + x1*x6*x7*x8*x9 + x5*x6*x7*x8*x9, 
x1*x2*x3*x4*x5*x6 + x2*x3*x4*x5*x6*x7 + x3*x4*x5*x6*x7*x8 +
x1*x2*x3*x4*x5*x9 + x1*x2*x3*x4*x8*x9 + x1*x2*x3*x7*x8*x9 +
x1*x2*x6*x7*x8*x9 + x1*x5*x6*x7*x8*x9 + x4*x5*x6*x7*x8*x9,
x1*x2*x3*x4*x5*x6*x7 + x2*x3*x4*x5*x6*x7*x8 + x1*x2*x3*x4*x5*x6*x9 + 
x1*x2*x3*x4*x5*x8*x9 + x1*x2*x3*x4*x7*x8*x9 + x1*x2*x3*x6*x7*x8*x9 + 
x1*x2*x5*x6*x7*x8*x9 + x1*x4*x5*x6*x7*x8*x9 + x3*x4*x5*x6*x7*x8*x9, 
x1*x2*x3*x4*x5*x6*x7*x8 + x1*x2*x3*x4*x5*x6*x7*x9 +
x1*x2*x3*x4*x5*x6*x8*x9 + x1*x2*x3*x4*x5*x7*x8*x9 + x1*x2*x3*x4*x6*x7*x8*x9 + 
x1*x2*x3*x5*x6*x7*x8*x9 + x1*x2*x4*x5*x6*x7*x8*x9 +
x1*x3*x4*x5*x6*x7*x8*x9 + x2*x3*x4*x5*x6*x7*x8*x9, x1*x2*x3*x4*x5*x6*x7*x8*x9 - 1]:;\\
proba\_epsilon:=1e-7;\\
debug\_infolevel:=1;\\
//gbasis\_reinject(-100);\\
threads:=36;\\
time(archive("H9",gbasis(cyclic9 ,indets(cyclic9))));\\
}\\
Uncommenting \verb|gbasis_reinject(-500);| will stop the computation
after 500 basis elements are reconstructed. To finish the
computation, run\\ 
\verb|gbasis(cyclic9,indets(cyclic9),gbasis_reinject=unarchive("H9"))|.

\subsection{How to compute cyclic10 over $\Q$}
The computation of cyclic10 was done in 2 stages on \verb|cocalc| by calling
\verb|icas| (Giac text interpreter) on the following script\\
{\tt \small
debug\_infolevel:=1;\\
threads:=36;\\
gbasis\_simult\_primes(12,800,9,1600,2); // 12 simult up to 800 primes, then 9, then 2\\
// gbasis\_reinject(.05,.05); \\
gbasis\_reinject(-2205); // so that we reinject the 2205 first basis elements after 796 primes\\
vars := [x0,x1,x2,x3,x4,x5,x6,x7,x8,x9];\\
sys := [x0 + x1 + x2 + x3 + x4 + x5 + x6 + x7 + x8 + x9,
x0*x1 + x0*x9 + x1*x2 + x2*x3 + x3*x4 + x4*x5 + x5*x6 + x6*x7 + x7*x8 + x8*x9,
x0*x1*x2 + x0*x1*x9 + x0*x8*x9 + x1*x2*x3 + x2*x3*x4 + x3*x4*x5 + x4*x5*x6 + x5*x6*x7 + x6*x7*x8 + x7*x8*x9,
x0*x1*x2*x3 + x0*x1*x2*x9 + x0*x1*x8*x9 + x0*x7*x8*x9 + x1*x2*x3*x4 + x2*x3*x4*x5 + x3*x4*x5*x6 + 
x4*x5*x6*x7 + x5*x6*x7*x8 + x6*x7*x8*x9,
x0*x1*x2*x3*x4 + x0*x1*x2*x3*x9 + x0*x1*x2*x8*x9 + x0*x1*x7*x8*x9 + x0*x6*x7*x8*x9 + 
x1*x2*x3*x4*x5 + x2*x3*x4*x5*x6 + x3*x4*x5*x6*x7 + x4*x5*x6*x7*x8 + x5*x6*x7*x8*x9,
x0*x1*x2*x3*x4*x5 + x0*x1*x2*x3*x4*x9 + x0*x1*x2*x3*x8*x9 + x0*x1*x2*x7*x8*x9 + x0*x1*x6*x7*x8*x9
 + x0*x5*x6*x7*x8*x9 + x1*x2*x3*x4*x5*x6 + x2*x3*x4*x5*x6*x7 + x3*x4*x5*x6*x7*x8 + x4*x5*x6*x7*x8*x9,
x0*x1*x2*x3*x4*x5*x6 + x0*x1*x2*x3*x4*x5*x9 + x0*x1*x2*x3*x4*x8*x9 + x0*x1*x2*x3*x7*x8*x9 + 
x0*x1*x2*x6*x7*x8*x9 + x0*x1*x5*x6*x7*x8*x9 + x0*x4*x5*x6*x7*x8*x9 + x1*x2*x3*x4*x5*x6*x7 + 
x2*x3*x4*x5*x6*x7*x8 + x3*x4*x5*x6*x7*x8*x9,
x0*x1*x2*x3*x4*x5*x6*x7 + x0*x1*x2*x3*x4*x5*x6*x9 + x0*x1*x2*x3*x4*x5*x8*x9 + 
x0*x1*x2*x3*x4*x7*x8*x9 + x0*x1*x2*x3*x6*x7*x8*x9 + x0*x1*x2*x5*x6*x7*x8*x9 + 
x0*x1*x4*x5*x6*x7*x8*x9 + x0*x3*x4*x5*x6*x7*x8*x9 + x1*x2*x3*x4*x5*x6*x7*x8 + x2*x3*x4*x5*x6*x7*x8*x9,
x0*x1*x2*x3*x4*x5*x6*x7*x8 + x0*x1*x2*x3*x4*x5*x6*x7*x9 + x0*x1*x2*x3*x4*x5*x6*x8*x9 + 
x0*x1*x2*x3*x4*x5*x7*x8*x9 + x0*x1*x2*x3*x4*x6*x7*x8*x9 + x0*x1*x2*x3*x5*x6*x7*x8*x9 + 
x0*x1*x2*x4*x5*x6*x7*x8*x9 + x0*x1*x3*x4*x5*x6*x7*x8*x9 + x0*x2*x3*x4*x5*x6*x7*x8*x9 + 
x1*x2*x3*x4*x5*x6*x7*x8*x9,x0*x1*x2*x3*x4*x5*x6*x7*x8*x9-1];\\
time(archive("H10\_2205",gbasis(sys ,vars)));
}\\
This computation ran in about one week on \verb|cocalc| 
(real time on \verb|cocalc| 0.68e6 seconds=7.9 days, CPU
time 16.7e6 seconds=193 days) and outputs the 2205 first generators of
the Groebner basis in the file \verb|H10_1|. I could also run with 64G of RAM in 2
weeks on \verb|ifnode2| with this configuration\\
\verb|threads:=24;|\\
\verb|gbasis_simult_primes(4,250,3,600,2);|\\
The output (compressed in gzip format) is available here\\
\verb|https://www-fourier.ujf-grenoble.fr/~parisse/giac/benchmarks/H10_2205.gz|

Now we can run \verb|icas| on a copy of the script above, after commenting the line\\
\verb|// gbasis_reinject(-2205)|,\\
uncommenting the line\\
\verb|gbasis_reinject(.05,.05);|,\\
and replacing the last line by\\
\verb|time(archive("H10",gbasis(sys,vars,gbasis_reinject=unarchive("H10_2205"))));|\\
The second stage runs in about 1.5 days and requires about 220G of
memory. Make sure you have enough disk
space to save the full basis (about 100G) or make a partial save, like
this~:\\
\verb|time(H:=gbasis(sys,vars,gbasis_reinject=unarchive("H10_2205")));|\\
\verb|archive("H10",eval(H,1)[0..100]):;|

Stage 2 may be split in several steps to save the temporary memory
required, for example \verb|gbasis_reinject(-3263)| will save
the 3263 first basis elements (this requires much less RAM, because we
compute about 1000 new basis elements instead of about 3000). 
Reconstruction in cyclic10 happens after the following numbers of primes \\
\begin{tabular}{|c|c|} \hline
Number of primes & Reconstructed basis elements \\ \hline
3 & 21 \\
388 & 41 \\
794 & 2205 \\
1041 & 2351 \\
1426 & 2636 \\
1933 & 3263 \\
2225 & 5690 \\
\hline
\end{tabular}

An attempt was made to re-inject the partial reconstruction of the 41 first
generators, but instead of speeding up the computation, it slowed it
down (more precisely, the 1st run with re-injection was about twice as
fast, but further runs were slower, as if learning was fully
ineffective). We observed the same behavior trying to decrease 
the maximal number of pairs with \verb|gbasis_max_pairs|, 
this make the 1st run faster, but after that learning is
much less effective and next runs are slower.

\section{Conclusion}
In 1999, J.C. Faug\`ere stated about the cyclic 9 computation on $\Q$
``This success is also a failure in some sense: the size of the output
is so big that we cannot do anything with this result''. I don't think
this is true anymore, the output of cyclic9 can be used
nowadays. Perhaps 20 years from now, the output of cyclic10 on $\Q$
will be used~?
% biblio: Buchberger, Faugere F4, Faugere F5, Allan Steel,
% Gebauer and Moller, Arnold, Joux and Vitse

\section*{Acknowledgements}
Many thanks to William Stein for letting me run most of these
computations on \\\verb|cocalc.sagemath.org|
 and to S. Leli\`evre for his interest.

\bibliography{gb.bib}

\end{document}